# Probability theory as a physical theory points to superdeterminism.

Louis Vervoort, PhD

*School of Advanced Studies,*

*University of Tyumen, Russian Federation*

06.08.2019

**Abstract**. Probability theory as a physical theory is, in a sense, the most general physics theory available, more encompassing than relativity theory and quantum mechanics, which comply with probability theory. Taking this simple fact seriously, I argue that probability theory points towards superdeterminism, a principle that underlies, notably, 't Hooft's Cellular Automaton Interpretation of quantum mechanics. Specifically, I argue that superdeterminism offers a solution for 1) Kolmogorov's problem of probabilistic dependence; 2) the interpretation of the Central Limit Theorem; and 3) Bell's theorem. Superdeterminism's competitor, indeterminism ('no hidden variables'), remains entirely silent regarding 1) and 2), and leaves 3) as an obstacle rather than a solution for the unification of quantum mechanics and general relativity. This suggests that, if one wishes to stick to the standard position in physics and adopt the principles with the highest explanatory power, one should adopt superdeterminism and reject indeterminism. Throughout the article precise questions to mathematicians are formulated to advance this research.

**1. Introduction**.

In this article I will explore what can be learnt from considering probability theory a physical theory, besides a mathematical one. I will start from a few straightforward hypotheses on probability as a physical property, and study some of their consequences. The main goal is to show that a coherent interpretation of probability offers an interesting lens to study several foundational problems. More specifically, I will argue that such an interpretation points towards determinism (the assumption that all events have causes[1] by which they follow with necessity) or more precisely superdeterminism (in short, the assumption that all events have causes tracing back to the Big-

---
[1] Such causes are mathematically represented by (hidden) variables.

Bang). Indeed, superdeterminism seems to have a stronger explanatory power than its competitor, indeterminism: it can coherently provide explanations for several foundational questions, which are left unanswered or unsolved by indeterminism. This result seems to go against the standard assumption that the world is ultimately indeterministic; superdeterminism is rarely adopted, but it underlies 't Hooft's Cellular Automaton Interpretation of quantum mechanics [1-3]. I will conclude that superdeterminism is a candidate as a general principle that could guide us in constructing new theories. For deriving this conclusion, as well as for a general introduction to superdeterminism, no complex mathematics is needed; but I will formulate precise questions to mathematicians to go further in this research. Finally, I will recall that a peaceful coexistence between the classic objective interpretation of probability and the subjective Bayesian interpretation, gaining in popularity, is possible under certain circumstances.

**2. Objective probability**.

As a mathematical theory, probability theory was axiomatized by Andrei Kolmogorov in his reference work of 1933 [4]. This work solved as a matter of principle all purely mathematical issues of probability theory. But probability theory is also a theory about the real world, in particular the physical world: it can be applied to physical situations. To do so, one needs to *interpret* the concept of probability. Interestingly, here is where subtleties and problems come in, giving rise to a long series of paradoxes (such as Bertrand's paradox, the Monty Hall problem, the Borel-Kolmogorov paradox, Bernstein's paradox etc.). As Tijms writes: "Probability theory is the branch of mathematics in which it is easiest to make mistakes" [5]. Many interpretations of probability are on the market, such as the classical interpretation of Laplace, the frequency interpretation, the subjective or Bayesian interpretation etc., as witnessed by a vast body of literature [6-11, 26]. In physics (and beyond), the standard interpretation is the frequency interpretation, stipulating, roughly, that an event's probability is the limit of its relative frequency in a large number of trials. (Note that this is a rough definition: phrased like this it is marvellous source for bad application of the theory – see further.) However, since about two decades a growing number of researchers in quantum information theory have become adepts of the subjective or Bayesian interpretation (see e.g. [51]), which posits that any probability value is a *subjective* degree of belief of an agent (e.g. [26]). It seems already undeniable that this interpretation has been a source of inspiration for new mathematical results in quantum information theory. At the same time it is undeniable that this subjective stance is orthogonal to the usually assumed objective character of physics, summarized by following fact: any well-defined probability density measured by Alice on a quantum system (by determining relative frequencies) can be reproduced at will by Bob (if he performs the same experiment). However, it is not difficult to show that any coherently defined subjective interpretation can be re-construed within an objective frequency interpretation; ultimately coherent versions of both interpretations are isomorphic[2] ([14], Ch. 5). Thus the inspirational advantage that some have found in Bayesianism does not need to be at odds with the classic assumption that will be adopted here, namely that probability values are objective *if measured in well-defined experiments* [10-14]. This is also in agreement with a recently derived result showing that both probability and essential features of quantum mechanics follow from very general (symmetry) principles, such as associativity, commutativity and closure: these are all objective in nature [19, 20].

The author who gave the most elaborated analysis of physical probability is, it seems, Richard von Mises[3] [10-11]. Based on his work it becomes clear, if it is not obvious from the start, that probability theory can also be seen as a *physical theory – for specific systems*. Many of von Mises' basic ideas are quite straightforward; I suppose they are intuitively, implicitly applied by most practitioners of probability theory. But it is in paradoxical or complex situations that an explicit formulation of a precise interpretation of probability becomes necessary, as argued below. In the following I will start by summarizing a few straightforward ideas that were elaborated in some detail in [12-14], mostly inspired by von Mises.

It seems clear that (objective) probability is a property that, strictly speaking, can only be attributed to *ensembles* of sufficiently similar systems, all characterized by a sufficiently similar 'environment' (or boundary conditions if one prefers). Usual physical properties can be measured on individual systems. But probabilities can only be measured by a series of sufficiently similar experiments (all involving sufficiently similar trial systems). Here it is important to remember that the precise boundary conditions of the experiment (the environment) determine the numerical values of the probabilities [12-14]. In this sense, probabilities 'emerge' out of (long runs of) experiments. The role of the environment or boundary conditions is often forgotten: in mundane cases no-one feels compelled to mention them. "The probability $P_6$ to throw a 6 with this normal die" seems well-defined; everyone expects $P_6 = 1/6$; and everyone knows how to verify this. And yet, as a matter of principle, $P_6$ is *not* well-defined; in principle one should consider and mention the boundary conditions of the die throw, and these conditions involve both 'initializing' and 'measurement' (or 'observing') conditions. The initializing conditions of a normal die throw involve randomizing (e.g. by sufficiently vigorous throwing), and the observing conditions involve e.g. interaction with a table. One may well imagine specific conditions of throwing a die on a table covered with glue, such that $P_6$, *in these experimental conditions*, is not = 1/6 but close to 1. Let's briefly look at a famous paradox, Bertrand's, that has intrigued a good number of famous probability theorists, including Kolmogorov [52]. It goes as follows: "A chord is drawn randomly in a circle. What is the probability that it is shorter than the side of the inscribed equilateral triangle?" The reader may try: after enough trying it becomes clear that three possible answers can be given, *depending on the physical realization of randomly choosing the chord* – this can be done in three non-equivalent ways[4]. Thus the problem is not well posed; one should specify the precise experimental conditions that allow to perform a well-defined probabilistic experiment, in particular the initializing conditions. Analyzing problems as the above confers the definite impression that probability paradoxes arise due to the neglect of the experimental conditions that determine any probability value[5]. The fact that even skilled mathematicians disagree about them, seems to point to this reminder: one should not forget that *applied* probability theory is a physical theory, and that probability as a physical property depends on the environment, the context – it is relational, as so many other physical properties (position, momentum, energy,…).

---

[2] In some more detail, in Ref. [14], Ch. 5 it is shown that Bayesianism à la Jaynes [26], if applied to physical experiments occurring in univocal conditions, can be translated into the frequency interpretation: the (background or prior) information or knowledge of the subjective interpretation corresponds to the precise experimental conditions of the objective interpretation.

[3] I will not consider nor use here von Mises' mathematical calculus of probability (the so-called 'calculus of collectives'): Kolmogorov's calculus is much simpler. So I focus here on von Mises' interpretational or physical theory. Kolmogorov refers to von Mises as the primary source for the physical interpretation of probability [4].

The role of the experimental boundary conditions becomes particularly clear in quantum mechanics. Quantum mechanics is a probabilistic theory in the sense that, in general, its measurable quantities are probabilistic. One of the key ideas of the orthodox Copenhagen interpretation is expressed in following quote by Bohr. It is taken from his 1935 reply to Einstein, Podolsky and Rosen in their debate on the completeness of quantum mechanics [15]: "The procedure of measurement has an essential influence on the conditions on which the very definition of the physical quantities in question rests". This resonates well with the idea that quantum properties and probabilities are determined by the observing conditions, or in other words the observing subsystem, of the experiment 'generating' them. But this dependence holds – in principle – for all probabilistic systems, quantum or classical. For classical systems one needs some attention to realize the influence of the observing system (recall the die example above); in quantum mechanics the 'in principle' becomes basic – as correctly emphasized by Bohr. As an example: the probability that an x-polarized photon passes a y-polarizer obviously depends on x and y. In this sense quantum properties are determined by the 'observer' or rather the 'observing system / conditions'. The enigmatic role of the 'observer' in quantum mechanics, who would collapse the wavefunction, is not different in classical physics (see [12]; in [19] an in essence identical conclusion is reached). *Any* probability value, quantum or classical, is determined by the 'observer' – or if one prefers the detector parameters. In the quantum realm, this property is often called 'contextuality', on which more in Section 3.

---

[4] One can for instance randomly chose two points (homogeneously distributed) on the circle by two independent spins of a pointer; a procedure that leads to the probability 1/3, as can be measured and calculated. But two other 'initializing conditions' (experimental ways of randomly determining a chord) exist that lead to a different probability.

[5] For probabilities 'out there in nature', the initializing and observing conditions are the natural environment.

Let us here also recall that the hallmark of probabilistic or random systems is 1) unpredictability of individual outcomes, and more importantly, 2) 'frequency stabilization' [10-14]. The genuine probabilistic nature of an event can, in the end, only be verified by the fact that its relative frequency, as it is measured in a repeated experiment, *stabilizes* towards a fixed number when the number of trials increases. All probabilistic systems, stemming from the enormous variety of probabilistic disciplines and sub-fields, show this frequency stabilization and satisfy the simple rules of probability theory. To the surely biased taste of this author, this 'universal scheme of necessity' has a mysterious touch to it. And much of the remainder of the article is devoted to exploring this aspect of physical probability.

**3. Superdeterminism, and the problems it solves**.

The above introductory remarks are intended to show that probability theory is also a physical theory. Note that it is moreover a vastly corroborated theory, describing an uncountable variety of types of systems (ensembles from chance games, fluid mechanics, population dynamics, quantum mechanics, engineering, biology, etc.). Therefore, in a sense, probability theory is the most general physical theory: quantum mechanics has to comply with it, but also general relativity, which is, as a deterministic theory, a special case of a probabilistic theory (with probabilities having values 0 or 1). This idea is in agreement with Ref. [19], showing that probability theory can be derived from very general symmetry principles ruling combination and concatenation.

Now, there is a long tradition in physics assuming that probabilities are deterministic deep down. This hypothesis traces back at least to Laplace, who famously considered probabilities as emerging from a deterministic substratum; a creature with infinite knowledge would not need the tool of probability but would be able to predict everything with necessity. Let us define 'determinism' as usual as the hypothesis that all (physical) events and systems are deterministic, i.e. that *all (physical) events have anterior causes that necessitate these events, making their probability 0 or 1*. Since physical events are either deterministic or probabilistic, determinism is in practice equivalent with following hypothesis:

> **HYP-1**. Probabilistic systems are, in reality, deterministic systems in disguise. Any probability emerges from underlying deterministic processes.

We can define superdeterminism, as encountered in discussions of Bell's theorem, slightly more specifically as the assumption that all events have causes *tracing back to the Big-Bang*. 'Causes' can mathematically be defined as variables, as will be recalled further. Note that for our present purposes, superdeterminism can be considered equivalent to determinism (or HYP-1), since if all events have causes preceding them in time, then these causes also have anterior causes, and so on until the Big-Bang.

There are of course some well-known arguments in favor of HYP-1. Indeed, countless probabilistic systems are known to be reducible to deterministic systems – at least in principle, not always in practice. Think of statistical properties of fluid mechanical systems; fluid mechanics is a deterministic theory. Or think of all the probabilistic systems that are simulated by computer programs (e.g. by Monte Carlo methods): such codes always use pseudo-random number generators that are actually based on deterministic (but random-looking) functions. Hence these countless simulated systems are examples of probabilistic systems that can be considered deterministic under the surface. This already suggests that probability theory is closely related to causality / determinism.

The most obvious argument in favor of HYP-1 is the following. Any practitioner of probability theory *intuitively* acknowledges the link between causality and probability, via the concept of probabilistic dependence. One knows that there is a (somewhat subtle) link between correlation and causation; note that this link has been verified countless times. Indeed, recently books have been published trying to mathematically describe various forms of causal structures with probabilistic tools (in essence, the concept of probabilistic dependence), e.g. [16]. Attempts have been made to generalize this approach to the quantum realm, e.g. [29]. But it seems rarely acknowledged that this is a genuine argument in favor of the causal / deterministic interpretation of probability theory, HYP-1. To put things in a slightly more dramatic perspective, here is a question by Kolmogorov in his *Foundations of the Theory of Probability* ([4], p. 9):

> "In consequence, one of the most important problems in the philosophy of the natural sciences is – in addition to the well-known one regarding the essence of the concept of probability itself – to make precise the premises which would make it possible to regard any given events as independent. This question, however, is beyond the scope of this book."

Now, HYP-1 allows answering Kolmogorov's question: two probabilistic events are independent if they are not causally connected, i.e. if one event is not causally determining the other one, and if there is no common cause determining the two events (this is often termed Reichenbach's principle, cf. [16, 29]). We intuitively feel that the outcomes of two thrown dice are independent because we intuitively feel that there is no causal connection between the outcomes – in normal circumstances. I retain: HYP-1 or the assumption of (super)determinism allows for a coherent explanation of stochastic dependence, while its rival, indeterminism, offers no explanation.

A further argument in favor of HYP-1 is given by the Central Limit Theorem (CLT) of probability theory. There is something intriguing about the CLT, to which contributed such eminent mathematicians as de Moivre, Laplace, Chebyshev, Lyapunov, Pòlya, Gnedenko and Kolmogorov. It has been termed the "unofficial sovereign of probability theory" ([5], p. 169, [17], p. 149). Here is what Sir Francis Galton thought about it ([18], p. 66):

> "I know of scarcely anything so apt to impress the imagination as the wonderful form of cosmic order expressed by the Law of Frequency of Error. The law would have been personified by the Greeks and deified, if they had known of it. It reigns with serenity and in complete self-effacement amidst the wildest confusion. The huger the mob, and the greater the apparent anarchy, the more perfect is its sway. It is the supreme law of Unreason. Whenever a large sample of chaotic elements are taken in hand and marshalled in the order of their magnitude, an unsuspected and most beautiful form of regularity proves to have been latent all along."

Are there good grounds why mathematicians become lyrical in the face of this theorem ? In essence, the CLT states that any probabilistic variable that is the average, i.e. the weighted sum, of many independent variables has a normal (Gaussian) distribution *independently* of the distribution of the contributing variables. Now, this theorem of pure mathematics is again interpreted *causally* in almost all works on (applied) probability. Textbooks typically present the theorem as an explanation of why the Gaussian distribution is so overwhelmingly present in the world, describing an unlimited variety of stochastic phenomena and properties. The textbook rationale is simple: many properties are the average result of many other independent phenomena; or more precisely: many properties are the *effect* of many independent *causes*. Hence we see deterministic reasoning appearing again in applied probability. This fact aligns with the conjecture that probability theory and the CLT are about the causal connectedness of things. Again, HYP-1 or (super)determinism allow for a coherent explanation of the CLT, whereas indeterminism remains silent.

But if this interpretation of the CLT is correct, then it is tempting to speculate that the theorem can be generalized. Note first that, in physics, the usual way to formalize causal dependence is via functional dependence: if a variable y is caused or determined by variables $(x_1,…,x_N)$ then $y = f(x_1,…,x_N)$ for some function f. Now, if it indeed suffices, as proposed in the usual rationale to interpret the CLT, that a variable is determined by a large number of causes to have a normal distribution, and if causal determination is expressed as functional dependence, then one is tempted to ask following question:

> **Q1**. Can the Central Limit Theorem be generalized as follows ?: "Under broad conditions, a stochastic variable that is a *function* of many independent variables will be normally distributed, independently of the distribution of the contributing variables".

This is the first of a series of questions related to HYP-1 (Q1 – Q5) addressed to mathematicians. The CLT seems to restrictively consider a special case for the function f, namely $y = \frac{1}{N}\sum_i x_i$. Here is a first hint. According to Taylor's theorem one can approximate a function $f(x_1,…,x_N)$ as follows:

$$f(x_1,…,x_N) = f(0,…,0) + [\,\partial f / \partial x_1\,]_0.x_1 + \ldots + [\,\partial f / \partial x_N\,]_0.x_N + O[x_i^2]. \qquad (1)$$

According to the CLT the dominant terms, linear in $(x_1,…,x_N)$, converge to a Gaussian.

Why, then, is HYP-1 not more popular, if there exist such cogent arguments in its favor? One reason is that the standard (Copenhagen) interpretation of quantum mechanics is in contradiction with it. Especially Niels Bohr but also Werner Heisenberg and other fathers of quantum theory have taught us that concepts as causality and determinism, ruling for centuries over science, are to be jettisoned. In other words, quantum mechanics would be *irreducibly* probabilistic and hidden causes are a heresy in the quantum world, contra HYP-1. Einstein, Podolsky and Rosen (EPR) attempted to contest this wisdom in 1935 in a famous paper, arguing for the incompleteness of quantum mechanics, but physicists usually assume that Bohr's reply won the debate [15]. However, this debate could not be proven or decided empirically; each of the debaters could well remain on his position. It is only in 1964 that John S. Bell devised an experiment that is generally believed to empirically decide the matter [21]. Bell's theorem ("local hidden-variable theories are in contradiction with quantum mechanics for certain experiments"), combined with the outcome of the Bell experiments, have convinced a majority of physicists that (local) hidden-variable theories, in particular deterministic ones, *do not exist*[6]. Hence HYP-1 would not work for quantum systems; quantum probabilities cannot be reduced to deterministic processes and theories in general; *Bohr wins again*. In conclusion, the only fundamental physical result against HYP-1 resides in Bell's theorem (and some related theorems). Let us therefore see on which assumptions this theorem is based; these are all probabilistic assumptions. Getting insight in Bell's theorem is useful for another reason: it seems to be a fundamental obstacle in the unification of quantum mechanics and general relativity.

In essence, Bell assumes that in a Bell experiment, in which a quantum property as spin or polarisation ($\sigma$) is measured on two entangled electrons or photons flying off in opposite directions, the average product of $\sigma_1$ and $\sigma_2$ is given by following formula, *if one assumes hidden determinism*:

$$M(a,b) \;=\; <\sigma_1(a).\sigma_2(b)> \;=\; \int \sigma_1(a,\lambda).\sigma_2(b,\lambda).\rho(\lambda).d\lambda. \qquad (2)$$

A formula as (2) is the standard expression for the average value of the product of two stochastic variables ($\sigma_1$, $\sigma_2$) that are determined by variables $\lambda$ having a probability distribution $\rho(\lambda)$ ($\lambda$ can be values taken at the time of emission of the particles from the source). So the $\lambda$ are the hidden variables that would restore a deterministic nature to the stochastic variables $\sigma_1$ and $\sigma_2$; 'a' and 'b' are the left and right polarizer directions along which $\sigma_1$ and $\sigma_2$ are measured. (Below also the slightly more general stochastic case (see Eq. (4)) will be considered, in which $\sigma_1$ and $\sigma_2$ are not functions of λ, but in which λ determines their probability, so that $P(\sigma_1|a,\lambda)$ and $P(\sigma_2|b,\lambda)$ are defined.) But (2) leads to a contradiction with experiment, while the quantum result for M(a,b) is vindicated. Hence an expression as $\sigma_1 = \sigma_1(a,\lambda)$ is generally considered a heresy; what makes sense in the quantum realm (what can be calculated and measured) are *irreducible probabilities* as $P(\sigma_1|a)$ or $P(\sigma_1,\sigma_2|a,b)$. Or so goes the standard interpretation.

---

[6] 'Local' means "involving only subluminal interactions", so non-local theories would violate relativity theory; therefore the adepts of nonlocal theories are rare. Hidden variables are hidden causes. Such variables can deterministically determine quantum properties or stochastically determine their probabilities, as explained further.

But here a few puzzling thoughts pop up immediately – or rather, a few questions to mathematicians. Formula (2) seems a priori a perfectly acceptable expression to physicists; it has been used and tested against experiment in countless other statistical contexts. But first note that (2) assumes 'partial' determinism, in that it still introduces a probability measure (ρ); HYP-1 stipulates that even this ρ should, in principle, be explainable by deterministic processes. (But for proponents of determinism this partial determinism would be a huge step forward, of course.) More to the point, note that a more general expression for the average product M(a,b) is:

$$M(a,b) = <\sigma_1(a).\sigma_2(b)> = \int \sigma_1(a,\lambda).\sigma_2(b,\lambda).\rho(\lambda|a,b).d\lambda, \qquad (3)$$

containing the conditional probability density ρ(λ|a,b) (note the agreement between Eq. (3) and the expression for the more general stochastic case, Eq. (4) below). In the most general case, ρ(λ|a,b) ≠ ρ(λ): λ and (a,b) are then correlated (this property is often termed 'measurement dependence', a form of contextuality [8, 22-24, 49]). For instance, in a static Bell experiment there can, at least in principle, exist long-range forces (as the electromagnetic and gravitational force or a yet unknown field) between the λ at the source and the polarizers. However, physicists have performed advanced *dynamic* Bell experiments that avoid, to all likelihood, this kind of delocalized interactions (one can impose that (λ, a, b) are mutually spacelike separated by rapid switching of analyzers between the values (a,a') and (b,b'); for recent references see e.g. [47-48]). So for these dynamic experiments it is generally believed that Eq. (2) is indeed the right expression. Again, it is this expression that leads to the conclusion that HYP-1 is false; if Eq. (3) is the correct expression Bell's theorem is invalid (as a physical theorem eliminating local hidden-variables, not of course as a mathematical theorem). Since the stakes are so high, we ask:

> **Q2**. In Eq. (2), should ρ(λ|a,b) be used instead of ρ(λ) to describe hidden-variable scenarios in Bell experiments (even in the most advanced dynamic ones)?

A more general question is:

> **Q3**. Are there mathematical conditions in which Eq. (2) is not applicable to calculate the average product of two stochastic variables $<\sigma_1.\sigma_2>$ that are determined by other variables $\lambda$ ?

It seems the answer to Q3 is 'yes' for instance if there are realistic deterministic scenarios in which no probability density $\rho(\lambda)$ exists; but as far as I know there are no compelling reasons to assume this. But in any case, if mathematical conditions stipulated in Q2-Q3 can be found and correspond to realistic physical conditions (which is clearly suggested by the fact that applied probability theory is a physical theory), then Bell's analysis is not applicable.

Let us focus on Q2. Could there be correlations between freely chosen analyzer settings (a and b) and spacelike separated particle properties ? A handful of researchers have proposed that the answer to Q2 is yes, by arguing that *in a truly deterministic world – and not just in a partially deterministic one as in (2) –* there can, in principle, exist correlations between the particles and the choices 'a' and 'b' (a list of references is given in [8], [22-24], [49]). This objection to Bell's theorem is old, Bell himself was aware of it; but it is usually dismissed as a 'conspiratorial' solution, in 'obvious' contradiction with the existence of free will. According to this prevailing position, $\rho(\lambda|a,b)$ or equivalently $\rho(a,b|\lambda)$ make no sense: freely chosen parameters cannot be stochastically determined by particle properties – this would contradict free will. But proponents of determinism and HYP-1 may well point to the metaphysical character of this argument: it does not follow from a physics theory. They argue that, in principle, since the Big Bang *all* events have common causes and are therefore in principle correlated – the argument of superdeterminism. Moreover, invoking 'free will' to justify (2), as is often done in the quantum foundations community, is neglecting a mainstream conclusion from other communities: the *majority* of researchers (e.g. neurobiologists, cognitive neuroscientists, philosophers) having studied free will professionally have come to the conclusion that free will is *compatible* with total determinism ([25], p. 242). According to these experts one cannot invoke free will to justify (2).

It seems that the strongest argument against superdeterminism is the following: it would be a mind-boggling cosmic conspiracy that $\rho(\lambda|a,b)$ would always – whatever the precise sequences of the choices of a and b are, and whatever the way a and b are determined (be it by free choice of experimenters or by the number of mouse droppings in a lab [2] or by cosmic photons [47-48]) – be exactly such that it gives rise (via (3)) to the correct quantum value M(a,b) = cos(a-b). This argument surely looks convincing from one point of view. But again opponents may object that it is an anthropocentric argument: what is mind-boggling to our limited minds, may be an obvious truth for a more developed intelligence. (Conspiracy is often in the eye of the beholder. Seeing a conspiracy by higher powers is, most of the time, not seeing the necessary deterministic processes under the surface, as was already argued convincingly by Spinoza, a philosopher who put determinism at the center of his worldview [24].)

Can we go further in this debate ? In Ref. [53] an experimental test of superdeterminism is proposed; below I propose more arguments, tests, and questions to further scrutinize superdeterminism.

Sure enough, proponents of superdeterminism will admit that (2) is a formula that has always worked for macroscopic systems. To the best of my knowledge, there is no exception known to (2) in the macroscopic world in comparable statistical cases: an average value of a stochastic variable $x(\lambda) = \sigma_1(a,\lambda).\sigma_2(b,\lambda)$ can be calculated by (2) by integrating over $\rho(\lambda)$, the latter density being independent of freely or randomly chosen (a,b). Still, it may make sense to perform experiments on well-chosen macroscopic systems to verify this claim. In Ref. [27] it is argued that fluid-mechanical pilot-wave systems are interesting candidates to answer Q2 or Q3 experimentally; they consist of oil droplets resonantly interacting with a surface wave. These fluid systems, with a 'wave-particle' nature, have an intriguing capacity to mimic quantum properties [40-43]. Furthermore they have a holistic aspect to them, in that they are determined by the boundary conditions of the whole experimental set-up (just as quantum systems); as a consequence they show massive correlation (again just as quantum systems)[7] [40-43]. This suggests that it would be interesting to perform a Bell-type experiment on these systems in order to test Eqs. (2) and (3) [27].

On a theoretical level, perhaps it is possible to find quantitative arguments for Eq. (3) by considering 'supercorrelated' model systems, in which all variables are determined and connected, or correlated. Such physical model systems are Ising spin-lattices, simple and well-known types of cellular automata, in which each spin is correlated to each other spin [24, 27]. These systems therefore share at least one essential feature of 't Hooft's Cellular Automaton Interpretation (CAI), namely the full correlation of all variables. When one defines a (static) Bell-type experiment on such spin-lattices, Eq. (3) rather than Eq. (2) holds and the Bell inequality can be violated, even if these systems are local in the sense of satisfying the Clauser-Horne factorability condition [27] (defined under Eq. (4)). It would be interesting to upgrade this system to a more realistic, dynamical one, allowing the simulation of dynamical Bell-tests. This can perhaps be done by using the framework developed by Elze in [44-46], adopting essential features of the CAI. Thus, it seems worthwhile to continue this line of research, and to investigate the question:

> **Q4**. Are there model systems (such as cellular automata, spin-lattices,…) that can help answer Q2-Q3 ? E.g. systems that are deterministic at bottom, and in which probabilities as $\rho(\lambda|a,b)$ emerge (even when ($\lambda$, a, b) are mutually spacelike separated).

---

[7] Finally, these are nonlinear hence potentially chaotic systems, as has been experimentally proven. That means that even infinitesimally weak (but non-zero) nonlocal effects in them might lead to violation of the Bell inequality [28].

(Note that spin-lattices are not deterministic in the narrow sense, but they are causal or 'hidden-variable' systems, and that is what counts: one can define and express the Bell-probabilities $P(\sigma_1,\sigma_2|a,b)$ as a sum over probabilities determined by additional spin variables.) 't Hooft's Cellular Automaton Interpretation of quantum mechanics is evidently much more sophisticated than e.g. spin-lattices, but it is a cellular automaton model too. Other related models are investigated in [33, 44-46]. That cellular automata are potent devices to simulate complex dynamical systems, is for instance shown by the fact they can reproduce the Navier-Stokes equation [34].

Returning to question Q2, here is what seems the most convincing semi-quantitative argument why Eq. (3) could be correct in the sub-quantum realm, while Eq. (2) works in the macroscopic realm. Roughly put, it is based on the idea that the $\lambda$ are not just any sub-quantum variables, but the 'ultimate' variables taken at the Big-Bang[8]. Indeed, assume for a moment that the $\lambda$ are the degrees of freedom arising in the Theory of Everything, as envisaged, notably, by 't Hooft [1-3]. In a nutshell, 't Hooft posits the existence of a basis of ultimate 'ontological states' (characterized by $\lambda$), with associated Hilbert space, a preferred basis of deterministic states of which all particles are built. The ontological states evolve by permutations among themselves in discrete time intervals as in a cellular automaton. 'Beables' are operators that are diagonal on this orthonormal set of ontological states; their eigenvalues describe physical properties. In 't Hooft's CAI quantum mechanics emerges as a tool to calculate and predict experimental results, rather than as a fundamental theory. Regarding the nature of the hidden DOF in his theory 't Hooft says: "This set consists of the states the universe can 'really' be in. At all times, the universe chooses one of these states to be in, with probability 1, while all others carry probability 0" ([2], p. 14). And more specifically on the correlations in ρ in Eq. (3):

> "How to explain this apparent 'conspiracy'? A state considered in some experimental setup may either be a physical state, which we shall call 'ontological', or it is a superposition of ontological states. […] However, if an 'ontological basis' exists, which we believe to be the case, then there is a conservation law: the ontological nature of a state is conserved in time. If, at some late time, a photon is observed to be in a given polarization state, just because it passed through a filter, then that is its ontological state, and the photon has been in that ontological state from the moment it was emitted by its source. […] Indeed, the same conclusion can be reached by considering the black hole microstates, which quite possibly correspond to the ultimate, classical degrees of freedom of an underlying theory, while they fundamentally arise at the Planck scale only" ([1], p. 13-14).

---

[8] They thus easily escape from the spectacular attempt to probe 600-year old variables correlated with light from Milky Way stars [47].

Thus, "conservation of ontology" implies that the conspiratorial-*looking* correlations of Eq. (3), originating at the Big-Bang, are conserved.

Generally speaking, going beyond the CAI model, if the λ are the 'ultimate' variables taken at the Big-Bang, one can certainly not exclude that at this very fundamental level "everything is causally connected or correlated with everything" (again, an idea that the determinist finds highly convincing, nay obvious), and that one has to consider, *for these special, ultimate degrees of freedom,* their correlations with all other variables, including with a and b. Or perhaps one should say: one cannot exclude this correlation *in existence theorems as Bell's*, even if there is no practical way to ever establish the numerical values of ρ(λ|a,b). Then Eq. (3) rather than Eq. (2) is the correct formula. Only when considering sets of 'usual' variables as ($\sigma_1$, $\sigma_2$, a, b) that belong to the macroscopic (a, b) or quantum ($\sigma_1$, $\sigma_2$) level can one sometimes, in practice, assume independence between variables (e.g. a and b are independent as free variables to any experimental precision, as one can easily verify). Independence of such variables is fine for all practical purposes; dependence cannot be retraced by finite experiments. But already on the macroscopic and quantum levels some independencies make no sense anymore (e.g. $P(\sigma_1, \sigma_2)$ does not make sense, one has to consider $P(\sigma_1, \sigma_2|a,b)$; more precisely, $P(\sigma_1, \sigma_2|a,b) \neq P(\sigma_1, \sigma_2)$). Going one stage deeper to the fundamental level λ then induces the aboriginal correlations as in (3). So in a nutshell, the idea is that at the macroscopic level correlations may be lost or washed-out due to a statistical or large-ensemble effect, while for the most fundamental degrees of freedom this is not the case – these keep their universal correlation; or perhaps more precisely, one has to consider this correlation in existence theorems as Bell's theorem. Note that we do not need to ever be able to derive an explicit formula for these correlations in practice; we are only assuming here that such correlations exist.

Note that probability theory allows that general variables (a, b, λ) are correlated as in (3), even if (a, b) are *not* correlated: this is an application of Bernstein's paradox [30, 4]; and see the explicit calculation in [2]. Bernstein's paradox states that a set of pair-wise independent variables may not be jointly independent (there is no independence between all sub-ensembles of variables) [4, 31]. This may seem counterintuitive at first sight, but interestingly, probability theorists have found that this is the rule rather than the exception [31]. (No wonder, the determinist thinks: everything is in principle connected to everything.) This may offer a further line of research:

**Q5**. Does Bernstein's paradox offer insights in Q2-Q3 ?

Let us end with a simple proof that Bell's theorem evaporates in the above-mentioned conditions. (The first proof that Eq. (3) can lead not only to violation of the Bell inequality but to the correct quantum expression for M(a,b) was given by Brans [32].) Below model assumes that the ultimate variables are fully determined, correlated with emerging variables (as a, b) and that non-trivial ($\neq$ 0, 1) probabilities can emerge at a higher level – which is HYP-1. As was realized by Bell soon after his seminal article of 1964, Eq. (2) can actually be generalized to the stochastic case:

$$\mathrm{M(a,b)} = \sum_{\sigma_1 \sigma_2} \sigma_1 . \sigma_2 P(\sigma_1, \sigma_2 | a, b) = \sum_{\sigma_1 \sigma_2} \sigma_1 . \sigma_2 \int P(\sigma_1 | \lambda, \mathrm{a}) . P(\sigma_2 | \lambda, \mathrm{b}) \rho(\lambda | \mathrm{a}, \mathrm{b}) . d\lambda \ . \qquad (4)$$

This expression is only based on standard rules of probability calculus and on the physical assumption that $P(\sigma_1, \sigma_2 | a, b, \lambda) = P(\sigma_1 | a, \lambda) . P(\sigma_2 | b, \lambda)$. This is the so-called Clauser-Horne factorability condition, and is generally accepted as a consequence of local causality (relativity theory), since (λ, a, b) are mutually spacelike separated in advanced experiments. Note that the deterministic case (3) is a special case of (4), for $P(\sigma_1 | a, \lambda) = \delta_{\sigma_1, \sigma_1(a, \lambda)}$ and $P(\sigma_2 | b, \lambda) = \delta_{\sigma_2, \sigma_2(b, \lambda)}$ ($\delta$ is the Kronecker-delta). If the ab-initio degrees of freedom are fully determined, $\lambda$, having very many components, takes one value, $\lambda_0$, and

$$\rho(\lambda|a,b) = \delta(\lambda - \lambda_0(a,b)). \tag{5}$$

So $\lambda_0(a,b)$ contains the (very many) values of the components of $\lambda$ that lead to the choice (a, b) and to $P(\sigma_1, \sigma_2|a, b, \lambda_0(a,b))$. Then (4) becomes

$$M(a,b) = \sum_{\sigma_1 \sigma_2} \sigma_1.\sigma_2 P(\sigma_1|\lambda_0(a,b),a).P(\sigma_2|\lambda_0(a,b),b). \qquad \square \tag{6}$$

This type of M(a,b) can easily lead to a violation of the Bell inequality, as is well-known and straightforward to prove (note that $P(\sigma_1|\lambda_0(a,b),a)$ is also a function of b). In conclusion, this is a simple proof that under HYP-1, (super)determinism, Bell's theorem is invalid. Note once more that Eq. (4) is the more general variant of Eq. (3) *and that it follows directly from probability theory* (and one locality assumption). In view of the general empirical power of probability theory, emphasized in this article, I believe that this suffices to make us cautious about Eq. (2). Another article proposing a general interpretation linking causality, probability theory and Bell's theorem, with different conclusions than mine, is [50].

Thus HYP-1 or (super)determinism seems of a singular heuristic import, suggesting several avenues of research. HYP-1 states that all phenomena, including quantum ones, have causes. Maintaining, as quantum orthodoxy has it, that $P(\sigma_1|a) = ½$ is determined by *nothing*, so that the measured spin of an electron assumes a value say +1 rather than -1 on the basis of *nothing*, is difficult to accept for the determinist. Of course, it may well be that it is ultimately impossible to decide whether the universe is fully deterministic or probabilistic; but this is not the most important point; what is essential is the question whether extra variables (as deterministic or probabilistic causes) are conceivable (cf. footnote 6). One might therefore prefer to use 'determinism' in a somewhat broader sense (term it 'causality' or the 'hidden-variable hypothesis'), and to replace in HYP-1 "Any probability emerges from underlying deterministic processes" by "Any probability emerges from underlying causal processes, i.e. variables determining this probability".

If the above interpretation is correct, Bell's theorem teaches us *not* that nature is non-local[9] in the sense of superluminal; nor that nature is irreducibly acausal, indeterministic; but rather that there is no intermediate λ-level (underneath the quantum level) in which (2) is valid. In the level underneath the quantum level (3) is valid; ergo nature is deterministic (causal) and superconnected (and holistic or nonlocal in *this* very different sense). For proponents of determinism the truly mind-boggling aspect about Bell's theorem is that it allows us to know something directly about the most fundamental level of reality (the degrees of freedom of the Theory of Everything) – namely that there is only this one hidden-variable level underneath quantum mechanics, the level at which these degrees of freedom are correlated to 'everything'. Importantly, this interpretation of Bell's theorem incites one *to keep on searching for this local hidden-variable theory*. Recall that this interpretation of Bell's theorem comes down to an existence hypothesis – a 'no no-go' result; it does *not* necessitate that one can practically derive an expression for ρ(λ|a,b) where a and b are my choices of settings.

Once this interpretation adopted, it is tempting to envisage further lines of research, even if they are speculative and against mainstream views (however, need it be recalled that it is essential for the vitality of physics that mainstream views are challenged ?). One of the pressing problems of physics is the unification of quantum mechanics (or QFTs) and general relativity, which seems to butt on principled problems [1-3]. From the very abstract point of view of probability theory, arguably the most general theory available, it may be that the problem is this: quantum mechanics is a probabilistic theory, while relativity a deterministic one. But HYP-1 suggests that this schism is not insurmountable, after all. HYP-1, interpreted in its most straightforward way, suggests following (non-mainstream) strategy for the unification program: one should start from deterministic theories that can incorporate gravity and see whether these can generate quantum mechanics or QFT (as an emergent or effective theory). There exist theoretical results that indicate that this path is not to be excluded on a priori grounds, even if there is a general consensus that it is of great mathematical difficulty. For instance, fluid mechanics comes to mind: its mathematics, in essence the Navier-Stokes equation, is surprisingly powerful; better understanding it is one of the seven 'millennium problems' of mathematics. Indeed, it is well-known that the Navier-Stokes equation can be rewritten so as to yield the 1-particle Schrödinger equation [35-37]. (Upon closer inspection, the probabilistic character of the latter equation comes in via the assumption of stochastically fluctuating 'elementary units' or 'singularities' within a fluid element [36].) More recently, it was shown by Unruh that the Navier-Stokes equation can be rewritten as the equation for a massless scalar field in a geometry with a Schwarzschild metric near the horizon of a black hole. The quantized motion of sound waves in a convergent fluid flow is an analog model of a quantum field in a classical gravitational field. This allowed to predict the existence of 'sonic black holes' emitting a phononic version of Hawking radiation [38]; a prediction that has been verified [39]. In sum, from the perspective developed in this article, it seems that theoretical developments as 't Hooft's CAI and perhaps even the more classically-inspired ones just mentioned, deserve more attention in the unification program.

---

[9] I find it disturbing that not only mass media but also professional physicists keep on promoting the idea that Bell-experiments have now *'proven'* that nonlocality and spooky actions exist – in the sense that "measuring something here instantly affects something there far away". This is an interpretation, not a proof within an accepted physics theory. See e.g. this New York Times article, but almost any popular and many semi-professional text will do: https://www.nytimes.com/2015/10/22/science/quantum-theory-experiment-said-to-prove-spooky-interactions.html (dated 21.10.2015, retrieved 05.01.2019).

**4. Conclusion**.

It was argued here that probability theory, if taken seriously as an objective physical theory, offers a unifying framework that allows to address several foundational questions. Specifically, it was argued that there is one hypothesis which offers an explanation for 1) Kolmogorov's (or Reichenbach's) problem of probabilistic dependence; 2) the interpretation of the Central Limit Theorem; and 3) Bell's theorem – namely HYP-1, in short, the hypothesis of (super)determinism or causality; in mathematical terms, the hidden-variable assumption. (Superdeterminism and determinism can be considered equivalent, in this context, as recalled in Section 3.) On the other hand, *in*determinism ('no hidden variables') remains entirely silent regarding 1) and 2), and leaves 3) as an obstacle rather than an argument for the unification program. As far as I know, there are no problems or questions that could be solved by indeterminism and not by (super)determinism. Since physics usually adopts the hypothesis or principle that coherently answers more questions, this suggests, against popular views, that (super)determinism should be privileged over indeterminism as a guiding principle in physics. A number of mathematical questions were asked to go further in this research.

**Acknowledgements**. I would like to greatly thank, for discussion, Henry E. Fischer as well as the organizers and participants of the 2019 QIRIF conference on Quantum Foundations in Vaxjo, Sweden, notably A. Khrennikov, T. Nieuwenhuizen, G. Jaeger, P. Grangier.